\documentclass[11pt]{article}

\usepackage{amsmath,amssymb,amsfonts}
\usepackage{graphicx}
\usepackage{geometry}
\usepackage{caption}
\usepackage{subcaption}
\usepackage{float}
\usepackage{placeins}
\usepackage{hyperref}
\usepackage{bm}
\usepackage{authblk}

\title{Delay-induced chimera transitions via mode selection 
in a multiplex FitzHugh–Nagumo network}
\author{Hui Wu }

\affil[1]{hwu@cau.edu\\ Department of Mathematics, Clark Atlanta University, Atlanta, GA, USA}

\date{}

\begin{document}

\maketitle

\begin{abstract}
Time delay can fundamentally reshape collective dynamics in multiplex excitable networks. 
Here we investigate delay-induced pattern transitions in a two-layer nonlocally coupled FitzHugh--Nagumo network with delayed inter-layer interactions. 
Combining a mode-selection analysis with numerical simulations, we show that the inter-layer delay acts as a control parameter capable of inducing transitions among fragmented incoherent states, chimera-like partial coherence, and coherent traveling-wave states.

A Fourier mode decomposition and linear stability analysis reveal that delayed coupling selectively destabilizes and suppresses spatial modes through a nontrivial delay-dependent mechanism, providing an analytical explanation for the observed transitions. 
Numerical simulations confirm that increasing delay drives the system through three distinct dynamical regimes: strongly irregular fragmented patterns at small delay, chimera-like coexistence of coherent and incoherent spatial domains at intermediate delay, and re-stabilized coherent wave propagation at larger delay.

Our results demonstrate that deterministic delay alone can serve as a mode-selective control mechanism for generating and suppressing chimera-like states in multiplex excitable networks, providing insight into delay-mediated pattern formation in nonlinear multilayer systems.
\end{abstract}

\noindent\textbf{
Time delay is a fundamental ingredient in many natural and engineered networks, where it can alter stability and generate unexpected collective behavior. 
Here we show that delayed inter-layer coupling in a multiplex FitzHugh--Nagumo network can induce transitions among coherent traveling waves, chimera-like states, and restored coherent patterns. 
Using numerical simulations together with a mode-selection interpretation, we identify delay as an effective control parameter governing the emergence and reorganization of spatiotemporal structures. 
These results provide insight into how delays can be used to control pattern formation in multilayer nonlinear dynamical systems.
}

\section{Introduction}

Time-delayed interactions arise naturally in complex dynamical systems due to finite signal transmission speeds, processing times, and intrinsic memory effects \cite{hale1993,michiels2007,erneux2009,stepan1989}. 
Such delays are not merely perturbations of instantaneous dynamics; rather, they fundamentally reshape system behavior by introducing additional temporal degrees of freedom. 
As a consequence, delayed systems may exhibit oscillatory instabilities, multistability, and rich pattern formation phenomena that do not occur in their delay-free counterparts. 
In networked settings, time delays are known to alter synchronization mechanisms and generate nontrivial spatiotemporal structures, making them essential for modeling realistic interactions in physical, biological, and technological systems.

Synchronization in networks of coupled oscillators has been a central topic in nonlinear science, with applications ranging from neural dynamics to power grids and social systems \cite{strogatz2000,arenas2008,kuramoto1984}. 
Beyond the classical dichotomy between fully coherent and incoherent states, attention has increasingly focused on intermediate regimes exhibiting partial synchronization. 
In particular, chimera states—characterized by the coexistence of coherent and incoherent subpopulations—have emerged as a paradigmatic example of symmetry-breaking in homogeneous networks \cite{abrams2004,panaggio2015}. 
Originally identified in phase oscillator models, chimera states have since been observed in a wide variety of systems, including excitable media, neuronal networks, and multilayer architectures, highlighting their robustness and universality.

Multiplex or multilayer networks provide a natural framework for describing systems with multiple types of interactions \cite{kivela2014,boccaletti2014}. 
In such systems, intra-layer coupling governs local interactions, while inter-layer coupling links corresponding nodes across layers. 
Recent studies have shown that multiplex structures can significantly modify synchronization patterns and may enhance or suppress chimera states depending on the coupling architecture \cite{majumdar2016,ghosh2016}. 
When time delays are incorporated into inter-layer interactions, the resulting dynamics become even richer and less understood.

The FitzHugh--Nagumo (FHN) model is a prototypical system for describing excitable dynamics and has been widely used to study neuronal activity and pattern formation \cite{fitzhugh1961,nagumo1962}. 
Networks of coupled FHN units exhibit a variety of spatiotemporal behaviors, including traveling waves, cluster synchronization, and chimera states \cite{omelchenko2015,panaggio2015}.. 
Most existing studies focus on noise-induced or parameter-mismatch-induced transitions; however, the role of deterministic delay as a control parameter in multilayer FHN systems remains relatively unexplored.

In this work, we study a two-layer nonlocally coupled FHN network with delayed inter-layer coupling. 
Our primary goal is to understand how the delay parameter influences the emergence and transition of spatial patterns. 
Through systematic numerical simulations, we observe that increasing the delay drives the system through three distinct regimes:
\begin{itemize}
    \item coherent traveling-wave states at small delay,
    \item chimera-like states with spatial incoherence at intermediate delay,
    \item re-stabilized coherent states at larger delay.
\end{itemize}
These transitions are consistently captured by snapshots, space-time plots, and mean phase velocity profiles.

To provide a theoretical explanation, we perform a linear stability analysis based on Fourier mode decomposition. 
We show that the interplay between nonlocal coupling and delayed inter-layer feedback leads to a delay-dependent stability condition of the form
\[
\lambda_q = \text{local dynamics} + \text{nonlocal coupling} + \sigma_c e^{-\lambda_q \tau},
\]
where different spatial modes \(q\) respond differently to the delay. 
This mechanism explains why intermediate delay values selectively destabilize certain modes, leading to partial incoherence, while larger delays restore coherence through effective averaging.

The main contribution of this paper is to demonstrate that deterministic inter-layer delay alone can generate chimera-like patterns in multiplex excitable networks, without requiring noise, heterogeneity, or specially designed coupling. 
Our results provide new insight into delay-induced pattern formation and may be relevant for understanding multilayer neural systems and other complex networks with delayed interactions.

For clarity, Fig.~\ref{fig:schematic} illustrates the multiplex coupling architecture underlying the model studied here. 
This network structure provides the basis for both the analytical mode-selection mechanism and the delay-induced chimera transitions reported below.

The remainder of the paper is organized as follows. 
In Section~2 we formulate the model and derive the mode-selection framework. 
Section~3 contains numerical evidence for delay-driven transitions among incoherent, chimera-like, and coherent states. 
Section~4 concludes the paper with a summary and outlook.

\section{Linear Mode Analysis and Delay-Induced Instability Mechanism}
\label{sec:analysis}

In this section, we provide a linear mode analysis to explain the delay-induced transition observed in the numerical simulations. Although chimera states are nonlinear spatiotemporal patterns and generally do not admit a closed-form analytical solution, the onset of spatial incoherence can be understood through the stability of spatial Fourier modes. In particular, the inter-layer delay modifies the effective feedback phase and may destabilize selected nonzero spatial modes, leading to localized loss of coherence.

We consider a two-layer multiplex FitzHugh--Nagumo network with nonlocal intra-layer coupling and delayed inter-layer interactions. For each layer $\ell=1,2$, the activator and inhibitor variables of node $i$ are denoted by $u_{\ell i}$ and $v_{\ell i}$, respectively. The model is given by
\begin{align}
\varepsilon \dot u_{1i}(t)
&=
u_{1i}(t)-\frac{u_{1i}^3(t)}{3}-v_{1i}(t)
+\frac{\sigma}{2R}\sum_{j=i-R}^{i+R}
\Big[
b_{uu}(u_{1j}(t)-u_{1i}(t))
+b_{uv}(v_{1j}(t)-v_{1i}(t))
\Big]  \notag\\
&\quad
+\sigma_c\big(u_{2i}(t-\tau)-u_{1i}(t)\big),
\label{eq:model_u1}
\\
\dot v_{1i}(t)
&=
u_{1i}(t)+a
+\frac{\sigma}{2R}\sum_{j=i-R}^{i+R}
\Big[
b_{vu}(u_{1j}(t)-u_{1i}(t))
+b_{vv}(v_{1j}(t)-v_{1i}(t))
\Big],
\label{eq:model_v1}
\\
\varepsilon \dot u_{2i}(t)
&=
u_{2i}(t)-\frac{u_{2i}^3(t)}{3}-v_{2i}(t)
+\frac{\sigma}{2R}\sum_{j=i-R}^{i+R}
\Big[
b_{uu}(u_{2j}(t)-u_{2i}(t))
+b_{uv}(v_{2j}(t)-v_{2i}(t))
\Big]  \notag\\
&\quad
+\sigma_c\big(u_{1i}(t-\tau)-u_{2i}(t)\big),
\label{eq:model_u2}
\\
\dot v_{2i}(t)
&=
u_{2i}(t)+a
+\frac{\sigma}{2R}\sum_{j=i-R}^{i+R}
\Big[
b_{vu}(u_{2j}(t)-u_{2i}(t))
+b_{vv}(v_{2j}(t)-v_{2i}(t))
\Big].
\label{eq:model_v2}
\end{align}

The intra-layer coupling is represented by the rotational coupling matrix
\begin{equation}
B=
\begin{pmatrix}
b_{uu} & b_{uv}\\
b_{vu} & b_{vv}
\end{pmatrix}
=
\begin{pmatrix}
\cos\phi & \sin\phi\\
-\sin\phi & \cos\phi
\end{pmatrix}.
\label{eq:rotation_matrix}
\end{equation}
In the numerical simulations, $\phi=\pi/2-0.1$, so that the coupling contains a strong cross-coupling component between the activator and inhibitor variables. This type of coupling is known to facilitate phase-shifted patterns and chimera-like states in nonlocally coupled FitzHugh--Nagumo networks.

In the numerical simulations, $\phi=\pi/2-0.1$, so that the coupling contains a strong cross-coupling component between the activator and inhibitor variables. This type of coupling is known to facilitate phase-shifted patterns and chimera-like states in nonlocally coupled FitzHugh--Nagumo networks.

\begin{figure}[!t]
\centering
\includegraphics[width=0.74\textwidth]{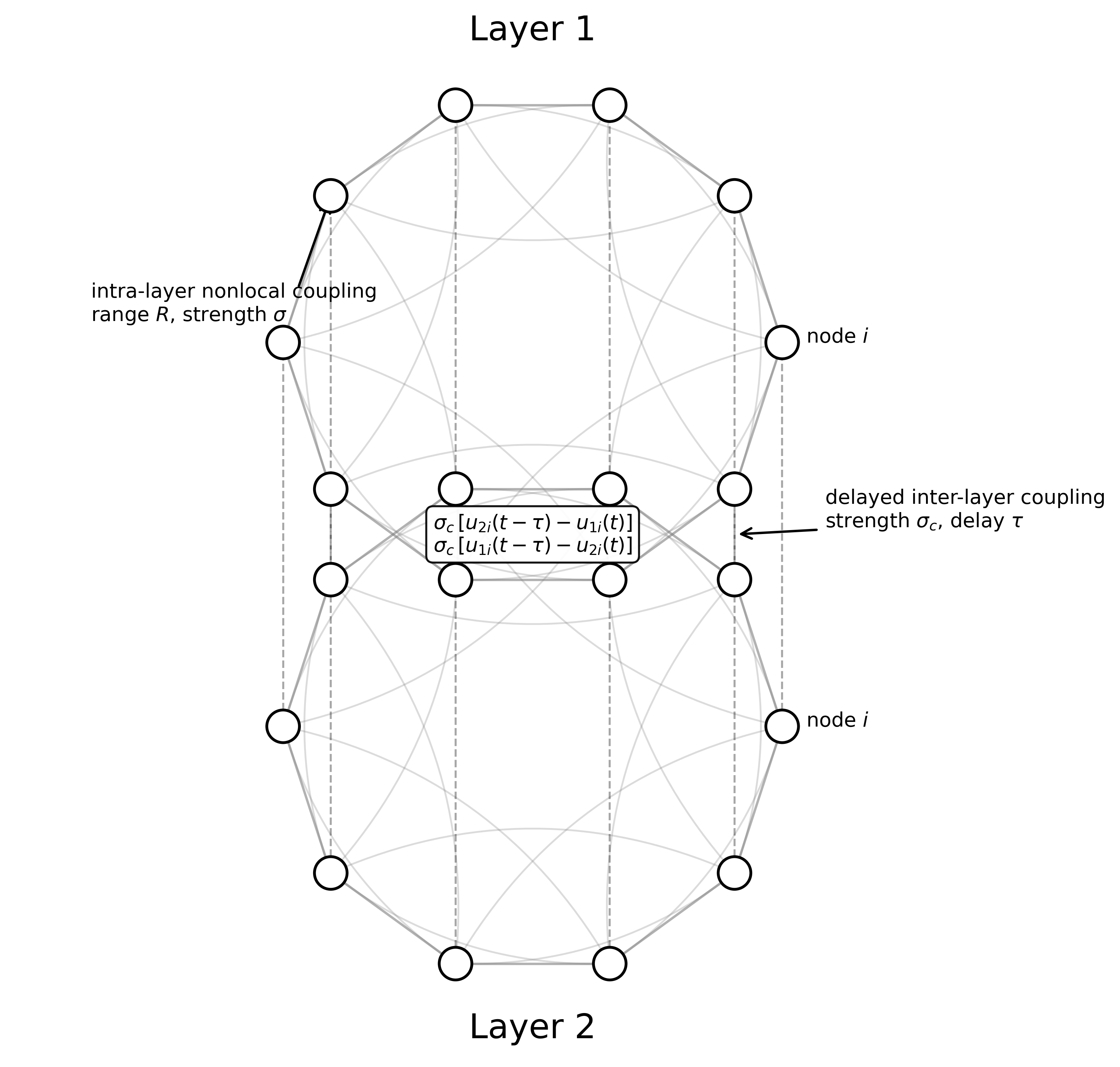}
\caption{
Schematic of the two-layer multiplex FitzHugh--Nagumo network.
Each layer consists of a nonlocally coupled ring with coupling range $R$ and intra-layer coupling strength $\sigma$.
Corresponding nodes across layers are coupled through delayed inter-layer interactions of strength $\sigma_c$ and delay $\tau$.
The inter-layer delay acts as the control parameter driving the mode-selection mechanism analyzed below.
}
\label{fig:schematic}
\end{figure}

Figure~\ref{fig:schematic} illustrates the multiplex coupling architecture and delayed inter-layer interactions considered in this work.

\subsection{Fourier decomposition of spatial perturbations}

Since each layer has a ring topology and the intra-layer coupling is translationally invariant, spatial perturbations can be decomposed into Fourier modes. We consider perturbations of the form
\begin{equation}
\delta u_{\ell i}(t)=\xi_{\ell}(t)e^{\mathrm{i}qi},
\qquad
\delta v_{\ell i}(t)=\eta_{\ell}(t)e^{\mathrm{i}qi},
\label{eq:fourier_perturbation}
\end{equation}
where $q$ is the spatial wave number. The mode $q=0$ corresponds to a spatially homogeneous perturbation, while $q\neq0$ corresponds to spatially structured perturbations.

For the nonlocal coupling term, we obtain
\begin{align}
\frac{1}{2R}\sum_{j=i-R}^{i+R}
\left(
\delta u_{\ell j}-\delta u_{\ell i}
\right)
&=
\frac{1}{2R}\sum_{m=-R}^{R}
\left(
e^{\mathrm{i}qm}-1
\right)
\xi_\ell e^{\mathrm{i}qi}.
\end{align}
Thus each Fourier mode is multiplied by the real factor
\begin{equation}
\Lambda_q
=
\frac{1}{2R}\sum_{m=-R}^{R}
\left(
e^{\mathrm{i}qm}-1
\right)
=
\frac{1}{R}\sum_{m=1}^{R}
\left(
\cos(qm)-1
\right).
\label{eq:lambda_q}
\end{equation}
The imaginary part cancels because the coupling range is symmetric. Notice that
\begin{equation}
\Lambda_0=0,
\qquad
\Lambda_q<0
\quad
\text{for }q\neq0.
\end{equation}
Therefore, the nonlocal coupling contributes a mode-dependent damping effect. For small $q$, using
\[
\cos(qm)-1\approx -\frac{q^2m^2}{2},
\]
we have the approximation
\begin{equation}
\Lambda_q
\approx
-\frac{q^2}{2R}\sum_{m=1}^{R}m^2
=
-\frac{q^2(R+1)(2R+1)}{12}.
\label{eq:lambda_small_q}
\end{equation}
This shows that long-wavelength modes are weakly damped, while nonzero spatial modes experience a damping strength depending on their wave number.

\subsection{Linearization around a coherent background}

Let $(u^*(t),v^*(t))$ denote a coherent background state. In the simplest local approximation, one may consider the frozen-time Jacobian of the FitzHugh--Nagumo vector field,
\begin{equation}
J_0(t)
=
\begin{pmatrix}
\dfrac{1-(u^*(t))^2}{\varepsilon} & -\dfrac{1}{\varepsilon}\\[2mm]
1 & 0
\end{pmatrix}.
\label{eq:J0}
\end{equation}
The intra-layer coupling contribution for the $q$th Fourier mode is
\begin{equation}
J_q^{\mathrm{intra}}
=
\begin{pmatrix}
\dfrac{\sigma\Lambda_q b_{uu}}{\varepsilon}
&
\dfrac{\sigma\Lambda_q b_{uv}}{\varepsilon}
\\[2mm]
\sigma\Lambda_q b_{vu}
&
\sigma\Lambda_q b_{vv}
\end{pmatrix}.
\label{eq:Jq_intra}
\end{equation}
Thus, even before introducing the inter-layer delay, the stability of each spatial mode is controlled by the product $\sigma\Lambda_q B$.

The delay enters only through the activator variable in the inter-layer coupling. Linearizing the delayed terms gives
\begin{equation}
\sigma_c\left(\delta u_{2i}(t-\tau)-\delta u_{1i}(t)\right)
\end{equation}
for layer 1, and
\begin{equation}
\sigma_c\left(\delta u_{1i}(t-\tau)-\delta u_{2i}(t)\right)
\end{equation}
for layer 2.

We now seek modal solutions of the form
\begin{equation}
(\xi_\ell(t),\eta_\ell(t))
=
(\xi_\ell,\eta_\ell)e^{\lambda t}.
\label{eq:lambda_ansatz}
\end{equation}
Then
\begin{equation}
\xi_\ell(t-\tau)
=
\xi_\ell e^{\lambda(t-\tau)}
=
e^{-\lambda\tau}\xi_\ell e^{\lambda t}.
\end{equation}
Therefore, the delay enters the characteristic equation through the exponential factor
\begin{equation}
e^{-\lambda\tau}.
\label{eq:delay_factor}
\end{equation}

\subsection{Symmetric and antisymmetric inter-layer perturbations}

Because the two layers are identical and coupled symmetrically, it is useful to decompose perturbations into symmetric and antisymmetric inter-layer modes.

For symmetric modes,
\begin{equation}
\xi_1=\xi_2,
\qquad
\eta_1=\eta_2.
\label{eq:symmetric_mode}
\end{equation}
The delayed inter-layer term becomes
\begin{equation}
\sigma_c\left(\xi(t-\tau)-\xi(t)\right)
=
\sigma_c\left(e^{-\lambda\tau}-1\right)\xi e^{\lambda t}.
\end{equation}
Therefore, the delayed feedback contribution for symmetric modes is
\begin{equation}
D_+(\lambda,\tau)
=
\begin{pmatrix}
\dfrac{\sigma_c}{\varepsilon}\left(e^{-\lambda\tau}-1\right) & 0\\[2mm]
0 & 0
\end{pmatrix}.
\label{eq:Dplus}
\end{equation}

For antisymmetric modes,
\begin{equation}
\xi_1=-\xi_2,
\qquad
\eta_1=-\eta_2.
\label{eq:antisymmetric_mode}
\end{equation}
Then
\begin{equation}
\xi_2(t-\tau)=-\xi_1(t-\tau),
\end{equation}
and the delayed inter-layer term becomes
\begin{equation}
\sigma_c\left(-\xi(t-\tau)-\xi(t)\right)
=
-\sigma_c\left(e^{-\lambda\tau}+1\right)\xi e^{\lambda t}.
\end{equation}
Hence the delayed feedback contribution for antisymmetric modes is
\begin{equation}
D_-(\lambda,\tau)
=
\begin{pmatrix}
-\dfrac{\sigma_c}{\varepsilon}\left(e^{-\lambda\tau}+1\right) & 0\\[2mm]
0 & 0
\end{pmatrix}.
\label{eq:Dminus}
\end{equation}

Combining the intrinsic dynamics, intra-layer coupling, and delayed inter-layer coupling, the characteristic equation for the $q$th spatial mode is
\begin{equation}
\det
\left[
\lambda I
-
J_0
-
J_q^{\mathrm{intra}}
-
D_\pm(\lambda,\tau)
\right]
=0.
\label{eq:characteristic_equation}
\end{equation}
Here, $D_+$ corresponds to symmetric inter-layer perturbations, while $D_-$ corresponds to antisymmetric inter-layer perturbations.

\subsection{Delay-induced non-monotonic stability}

The essential role of the delay can be seen by considering the crossing of characteristic roots through the imaginary axis. Let
\begin{equation}
\lambda=i\omega.
\end{equation}
Then the delay factor becomes
\begin{equation}
e^{-i\omega\tau}
=
\cos(\omega\tau)-i\sin(\omega\tau).
\label{eq:e_iomega}
\end{equation}
Thus the delay changes both the effective damping and the effective phase of the feedback.

For the symmetric mode, the delayed contribution is
\begin{align}
\sigma_c\left(e^{-i\omega\tau}-1\right)
&=
\sigma_c\left(\cos(\omega\tau)-1\right)
-
i\sigma_c\sin(\omega\tau).
\label{eq:symmetric_delay_complex}
\end{align}
Its real part is
\begin{equation}
\sigma_c\left(\cos(\omega\tau)-1\right)\le 0.
\end{equation}
Thus, in the symmetric sector, the delay often acts as a damping mechanism, but its imaginary part changes the phase relation among oscillators.

For the antisymmetric mode, the delayed contribution is
\begin{align}
-\sigma_c\left(e^{-i\omega\tau}+1\right)
&=
-\sigma_c\left(\cos(\omega\tau)+1\right)
+
i\sigma_c\sin(\omega\tau).
\label{eq:antisymmetric_delay_complex}
\end{align}
Its real part is
\begin{equation}
-\sigma_c\left(\cos(\omega\tau)+1\right)\le 0.
\end{equation}
However, the imaginary part again modifies the phase feedback. Since the stability condition depends on the full matrix in \eqref{eq:characteristic_equation}, including the intrinsic FitzHugh--Nagumo dynamics and the cross-coupled nonlocal interaction, the delay can destabilize selected spatial modes through a resonance-like mechanism.

The important point is that the stability of each spatial mode depends on
\begin{equation}
\cos(\omega\tau)
\quad\text{and}\quad
\sin(\omega\tau),
\end{equation}
and therefore varies non-monotonically with $\tau$. Consequently, increasing $\tau$ does not simply make the system more unstable. Instead, certain delay values destabilize selected spatial modes, while other delay values re-stabilize them.

\subsection{Interpretation of the numerical regimes}

The numerical simulations show that varying the delay parameter $\tau$ leads to a transition between spatially incoherent and coherent regimes. This behavior can be understood through the mode-dependent growth rates obtained from \eqref{eq:characteristic_equation}.

For a coherent traveling-wave state to remain stable, all nonzero spatial modes must be damped:
\begin{equation}
\operatorname{Re}\lambda_q(\tau)<0,
\qquad
q\neq0.
\label{eq:stable_condition}
\end{equation}
A chimera-like state emerges when the homogeneous or coherent background remains partially stable, but selected nonzero spatial modes become weakly unstable or nearly neutrally stable:
\begin{equation}
\operatorname{Re}\lambda_{q^*}(\tau)\approx0
\quad
\text{or}
\quad
\operatorname{Re}\lambda_{q^*}(\tau)>0
\end{equation}
for some $q^*\neq0$, while other modes remain damped. This creates a situation where some spatial regions remain coherent, while others develop persistent local desynchronization.

In the refined delay scan, three regimes are observed:

\paragraph{Case 1: $\tau=3.5$.}
For $\tau=3.5$, the space-time plot exhibits several fragmented bands and local incoherent patches. The corresponding mean phase velocity profile contains multiple plateaus and abrupt jumps. This suggests that several nonzero spatial modes are weakly unstable:
\begin{equation}
\exists q_1,q_2,\ldots
\quad
\text{such that}
\quad
\operatorname{Re}\lambda(q_j,3.5)>0.
\end{equation}
As a result, local desynchronization appears in multiple spatial regions, producing a strongly chimera-like or fragmented incoherent pattern.

\paragraph{Case 2: $\tau=4.0$.}
For $\tau=4.0$, the system displays a mixed regime. The space-time plot shows localized irregular structures embedded in a more coherent background. The mean phase velocity profile has a clear coexistence structure: part of the profile remains nearly flat, while another part exhibits fluctuations. This indicates that the system is close to a mode instability threshold:
\begin{equation}
\max_{q\neq0}\operatorname{Re}\lambda(q,4.0)\approx0.
\end{equation}
Therefore, the system supports coexistence of coherent and incoherent domains. This is the characteristic signature of a chimera-like state.

\paragraph{Case 3: $\tau=4.5$.}
For $\tau=4.5$, the space-time plot becomes regular and the mean phase velocity profile is nearly flat. This indicates that the previously unstable spatial modes have been re-stabilized:
\begin{equation}
\operatorname{Re}\lambda(q,4.5)<0
\quad
\text{for most or all }q\neq0.
\end{equation}
The system therefore returns to a coherent traveling-wave regime.

These observations demonstrate that the delay acts as a mode-selective control parameter. It can destabilize nonzero spatial modes at intermediate delay values, producing partial incoherence, and then re-stabilize those modes for larger delay values.

\subsection{Mean phase velocity as a diagnostic of coherence}

To quantify the spatial coherence of the network, we compute the mean phase velocity of each node. In the numerical simulations, this is approximated by counting the number of upward crossings of the activator variable:
\begin{equation}
\omega_i
=
\frac{2\pi M_i}{T},
\label{eq:mean_phase_velocity}
\end{equation}
where $M_i$ is the number of oscillation cycles of node $i$ over the observation time window $T$.

In a coherent traveling-wave state, all nodes share nearly the same mean phase velocity:
\begin{equation}
\omega_i\approx \omega_0
\qquad
\text{for all } i.
\end{equation}
In contrast, a chimera-like state is characterized by coexistence of coherent and incoherent domains:
\begin{equation}
\omega_i
=
\begin{cases}
\omega_0, & i\in \Omega_{\mathrm{coh}},\\
\text{spatially varying}, & i\in \Omega_{\mathrm{inc}},
\end{cases}
\label{eq:omega_chimera}
\end{equation}
where $\Omega_{\mathrm{coh}}$ and $\Omega_{\mathrm{inc}}$ denote coherent and incoherent spatial regions, respectively.

This diagnostic confirms the delay-induced transition observed in the simulations. For $\tau=3.5$, the mean phase velocity profile is highly fragmented, consistent with strong spatial incoherence. For $\tau=4.0$, the profile shows coexistence of flat and fluctuating regions, indicating a chimera-like state. For $\tau=4.5$, the profile becomes nearly flat, confirming the recovery of coherent traveling-wave synchronization.

\subsection{Summary of the mechanism}

The above analysis suggests that the observed delay-induced transition arises from the competition of three effects:

\begin{enumerate}
    \item The FitzHugh--Nagumo dynamics provides fast-slow excitable oscillations, making the system sensitive to phase and amplitude perturbations.
    \item The nonlocal intra-layer coupling introduces mode-dependent spatial damping through the factor $\Lambda_q$.
    \item The delayed inter-layer coupling introduces the factor $e^{-\lambda\tau}$, which changes the effective phase and damping of each spatial mode.
\end{enumerate}

As a result, the delay parameter $\tau$ selectively changes the stability of different spatial modes. Intermediate delay values can destabilize selected nonzero modes, resulting in local desynchronization and chimera-like patterns. Larger delay values may re-stabilize these modes, leading to the recovery of coherent traveling waves.

\subsection{Delay-induced mode selection mechanism}

To understand the delay-induced transition observed in the numerical simulations, 
we examine the effect of inter-layer delay on the stability of spatial modes.

Linearizing the system around a spatially coherent state and expanding perturbations 
into Fourier modes, each mode $q$ evolves with a characteristic growth rate $\lambda_q$ 
satisfying a delay-dependent characteristic equation of the form
\[
\lambda_q = A_q + B_q e^{-\lambda_q \tau},
\]
where $A_q$ represents the contribution from intra-layer nonlocal coupling, 
and $B_q$ arises from delayed inter-layer interactions.

The exponential term $e^{-\lambda_q \tau}$ plays a crucial role in shaping the stability landscape. 
In particular, it introduces a non-monotonic dependence of the growth rate on the delay $\tau$, 
effectively acting as a mode-selective filter.

For small delay values, the exponential term is close to unity, and the system behavior 
is dominated by the intrinsic spatial coupling. In this regime, unstable modes 
span a broad range of wave numbers, leading to fragmented incoherent dynamics.

As the delay increases, the exponential term selectively suppresses some modes 
while allowing others to remain weakly unstable. This partial stabilization leads to 
the coexistence of coherent and incoherent spatial regions, giving rise to chimera-like states.

For larger delay values, the exponential factor significantly damps all nontrivial modes, 
leaving only the dominant coherent mode stable. As a result, the system re-stabilizes 
into a coherent traveling-wave state.

Therefore, the delay parameter $\tau$ acts as a control mechanism that reshapes the 
modal stability spectrum, driving the system through a transition from incoherence 
to partial coherence and finally to full coherence.

This mechanism explains the numerically observed transition shown in Fig.~\ref{fig:main}.In particular, the first nontrivial Fourier mode plays a dominant role 
in determining the onset of the chimera-like regime.

\section{Numerical results}

We perform numerical simulations of the two-layer multiplex FitzHugh--Nagumo network 
to investigate the influence of the inter-layer delay parameter $\tau$ on spatiotemporal dynamics.

\begin{figure}[H]
\centering
\includegraphics[width=\textwidth]{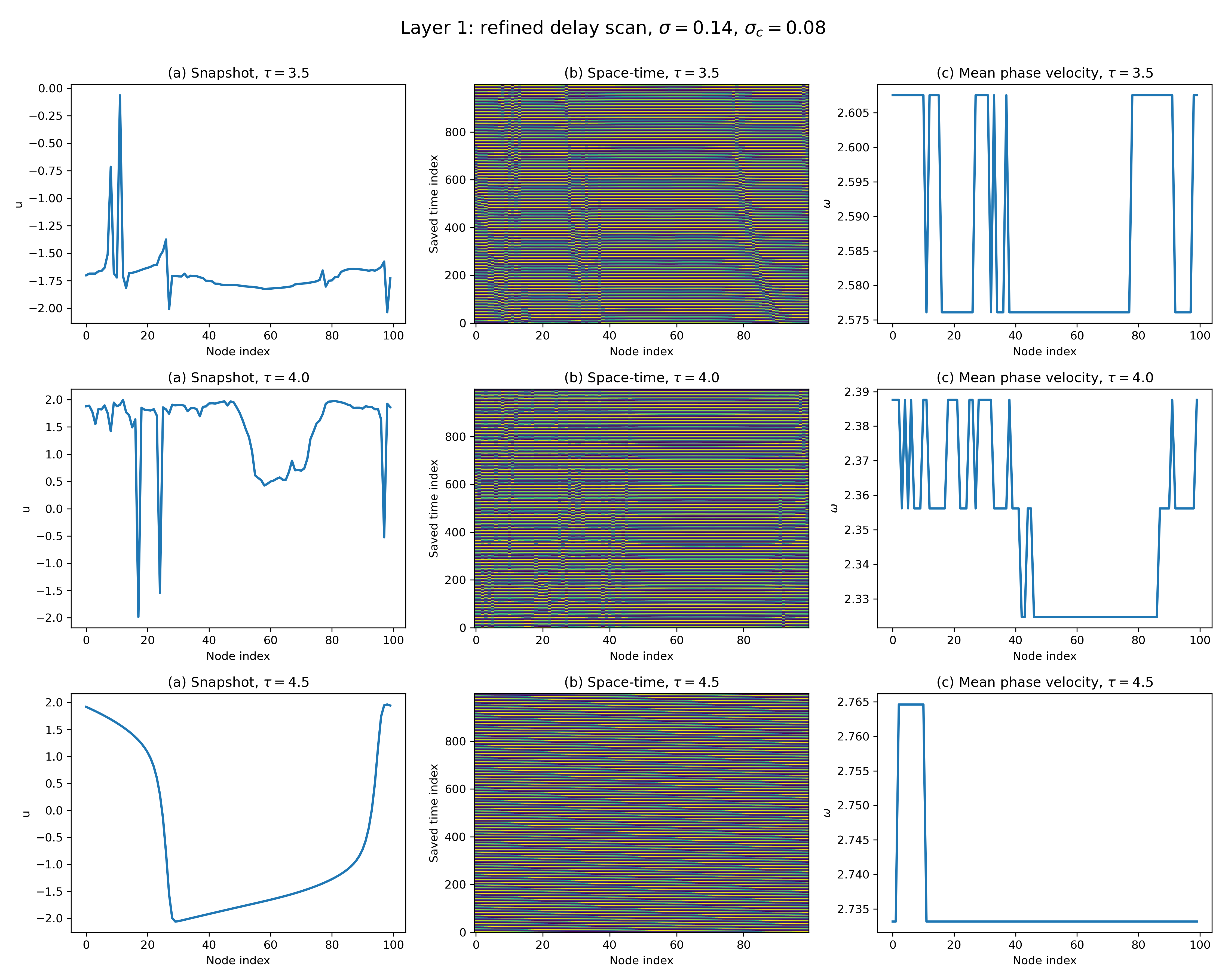}
\caption{
Spatiotemporal patterns for increasing delay values $\tau$.
Each row corresponds to a fixed delay ($\tau=3.5, 4.0, 4.5$ from top to bottom), 
while the columns show (a) snapshots of the activator variable, 
(b) space-time plots, and (c) mean phase velocity profiles $\omega_i$.
The parameters are fixed at $\sigma=0.14$ and $\sigma_c=0.08$.
}
\label{fig:main}
\end{figure}

Figure~\ref{fig:main} summarizes the main numerical findings. 
For representative delay values $\tau = 3.5, 4.0, 4.5$, we display 
(a) spatial snapshots, (b) space-time plots, and (c) mean phase velocity profiles.

\subsection{Delay-induced transition}

The selected delay values are obtained from a continuous parameter scan and represent typical dynamical regimes.

As the delay increases, the system undergoes a clear transition:
\[
\text{fragmented incoherence}
\;\longrightarrow\;
\text{chimera-like partial coherence}
\;\longrightarrow\;
\text{coherent traveling wave}.
\]

For $\tau = 3.5$, the system exhibits strongly irregular spatial structure. 
The snapshot shows fragmented incoherence, and the space-time plot contains disordered bands, 
indicating desynchronized dynamics. The mean phase velocity profile is highly nonuniform, 
confirming the absence of frequency synchronization.

At $\tau = 4.0$, coherent and incoherent regions coexist. 
The snapshot reveals localized distortions embedded in a more coherent background, 
while the space-time plot displays both regular and irregular structures. 
The mean phase velocity profile contains nearly flat regions coexisting with fluctuating segments, 
which is the characteristic signature of a chimera-like state.

For $\tau = 4.5$, the system returns to a coherent regime. 
The snapshot becomes smooth, the space-time plot shows regular stripe patterns, 
and the mean phase velocity profile is nearly flat, indicating that all oscillators 
share approximately the same frequency.

\subsection{Consistency across observables}

The transition is consistently observed across all diagnostics shown in Fig.~\ref{fig:main}. 
Snapshots capture instantaneous spatial structure, 
space-time plots reveal temporal evolution, 
and mean phase velocity profiles provide a quantitative measure of synchronization.

This consistency confirms that the observed regimes are robust and not transient artifacts.

\subsection{Interpretation}

The key observation is that deterministic inter-layer delay alone can generate and control 
spatial coherence patterns in a multiplex excitable network. 
Intermediate delay values destabilize uniform synchronization and induce mode competition, 
leading to partial coherence and chimera-like states, 
while larger delays suppress incoherence and restore coherent wave propagation.

\section{Conclusion}
\label{sec:conclusion}

We studied delay-induced collective dynamics in a two-layer multiplex FitzHugh--Nagumo network with nonlocal intra-layer coupling and delayed inter-layer interactions. The model is deterministic; no noise or parameter heterogeneity is introduced. Our goal was to understand whether inter-layer delay alone can act as a control mechanism for spatial coherence and chimera-like pattern formation.

Numerical simulations show that the delay parameter produces a clear transition among distinct dynamical regimes. For smaller delay values, the system exhibits fragmented incoherent patterns. At an intermediate delay value, coherent and incoherent spatial regions coexist, producing a chimera-like state. For larger delay, the system re-stabilizes into a coherent traveling-wave regime. These regimes are consistently identified through spatial snapshots, space-time plots, and mean phase velocity profiles.

To explain this behavior, we developed a linear Fourier mode analysis. The nonlocal coupling introduces a mode-dependent spatial factor \(\Lambda_q\), while the delayed inter-layer coupling contributes the exponential term \(e^{-\lambda\tau}\) to the characteristic equation. As a result, the stability of spatial modes depends non-monotonically on the delay through \(\cos(\omega\tau)\) and \(\sin(\omega\tau)\). This provides a mechanism by which intermediate delays can destabilize selected nonzero spatial modes, whereas larger delays can re-stabilize them.

The main conclusion is that deterministic inter-layer delay can serve as a mode-selective mechanism for inducing and suppressing chimera-like states in multiplex excitable networks. This suggests that delay is not merely a destabilizing factor, but an effective control parameter for organizing spatiotemporal coherence in multilayer systems. Future work may consider asymmetric delays, heterogeneous layers, and the interaction between deterministic delay and stochastic perturbations.

\bibliographystyle{unsrt}
\bibliography{references}

@book{hale1993,
  author={Hale, J. K. and Verduyn Lunel, S. M.},
  title={Introduction to Functional Differential Equations},
  publisher={Springer},
  year={1993}
}

@book{michiels2007,
  author={Michiels, W. and Niculescu, S.-I.},
  title={Stability and Stabilization of Time-Delay Systems},
  publisher={SIAM},
  year={2007}
}

@book{erneux2009,
  author={Erneux, T.},
  title={Applied Delay Differential Equations},
  publisher={Springer},
  year={2009}
}

@book{stepan1989,
  author={Stepan, G.},
  title={Retarded Dynamical Systems},
  publisher={Longman},
  year={1989}
}

@article{strogatz2000,
  author={Strogatz, S. H.},
  title={From Kuramoto to Crawford: exploring synchronization},
  journal={Physica D},
  volume={143},
  pages={1--20},
  year={2000}
}

@article{arenas2008,
  author={Arenas, A. and Diaz-Guilera, A. and Kurths, J. and Moreno, Y. and Zhou, C.},
  title={Synchronization in complex networks},
  journal={Physics Reports},
  volume={469},
  pages={93--153},
  year={2008}
}

@book{kuramoto1984,
  author={Kuramoto, Y.},
  title={Chemical Oscillations, Waves, and Turbulence},
  publisher={Springer},
  year={1984}
}

@article{abrams2004,
  author={Abrams, D. M. and Strogatz, S. H.},
  title={Chimera states for coupled oscillators},
  journal={Phys. Rev. Lett.},
  volume={93},
  pages={174102},
  year={2004}
}

@article{panaggio2015,
  author={Panaggio, M. J. and Abrams, D. M.},
  title={Chimera states: coexistence of coherence and incoherence},
  journal={Nonlinearity},
  volume={28},
  pages={R67--R87},
  year={2015}
}

@article{kivela2014,
  author={Kivel{\"a}, M. and others},
  title={Multilayer networks},
  journal={J. Complex Networks},
  volume={2},
  pages={203--271},
  year={2014}
}

@article{boccaletti2014,
  author={Boccaletti, S. and others},
  title={The structure and dynamics of multilayer networks},
  journal={Physics Reports},
  volume={544},
  pages={1--122},
  year={2014}
}

@article{ghosh2016,
  author={Ghosh, S. and Banerjee, T. and Dana, S. K.},
  title={Chimera states in multiplex networks},
  journal={Phys. Rev. E},
  volume={94},
  pages={032215},
  year={2016}
}

@article{majumdar2016,
  author={Majumdar, A. and Ghosh, D.},
  title={Chimera states in multiplex networks},
  journal={Chaos},
  volume={26},
  pages={103107},
  year={2016}
}

@article{fitzhugh1961,
  author={FitzHugh, R.},
  title={Impulses and physiological states},
  journal={Biophysical Journal},
  volume={1},
  pages={445--466},
  year={1961}
}

@article{nagumo1962,
  author={Nagumo, J. and Arimoto, S. and Yoshizawa, S.},
  title={An active pulse transmission line},
  journal={Proc. IRE},
  volume={50},
  pages={2061--2070},
  year={1962}
}

@article{omelchenko2015,
  author={Omelchenko, I. and Provata, A. and H{\"o}vel, P. and Sch{\"o}ll, E.},
  title={Robustness of chimera states in FHN networks},
  journal={Phys. Rev. E},
  volume={91},
  pages={022917},
  year={2015}
}

\end{document}